# Controlling complex policy problems: a multimethodological approach using system dynamics and network controllability

Lukas Schoenenberger[a] and Radu Tanase[b]

[a]Department of Business, Health, and Social Work, Bern University of Applied Sciences, Bern, Switzerland
[b]URPP Social Networks, University of Zurich, Zurich, Switzerland

Abstract

*Notwithstanding the usefulness of system dynamics in analyzing complex policy problems, policy design is far from straightforward and in many instances trial-and-error driven. To address this challenge, we propose to combine system dynamics with network controllability, an emerging field in network science, to facilitate the detection of effective leverage points in system dynamics models and thus to support the design of influential policies.*

*We illustrate our approach by analyzing a classic system dynamics model: the World Dynamics model. We show that it is enough to control only 53% of the variables to steer the entire system to an arbitrary final state. We further rank all variables according to their importance in controlling the system and we validate our approach by showing that high ranked variables have a significantly larger impact on the system behavior compared to low ranked variables.*

# Introduction

System dynamics (SD), an approach to modeling and simulating complex systems, has repeatedly demonstrated its value in contributing to the understanding and solution of complex policy problems—most notably in areas such as public health, energy and the environment, social welfare, sustainable development and security (Ghaffarzadegan *et al.*, 2010; Sterman, 2000). Particularly in large (complex) SD models, however, the detection of model levers, i.e., variables capable of effectively and efficiently controlling complex policy problems, is a challenge. This is due to the high degree of interdependent model variables and nonlinear relationships typically present within these models. So, notwithstanding the usefulness of SD in the analysis of complex policy problems, the solution identification process (policy design) is far from trivial and in most cases trial-and-error driven (Forrester, 1994; Oliva, 2016). To address this challenge, we propose a multimethodological approach combining SD with network controllability to enhance the speed and quality of model lever discovery in SD models. In this respect, our article is a first attempt to bring SD a step closer to a very recent and fast growing field with strong roots in complex systems research: network science; thereby abiding by Anderson's (2014) and Barlas' (2016) call to reach out and partner with emerging systemic disciplines.

In their first editorial in *Network Science*, i.e., a novel journal published by Cambridge University Press, Brandes *et al.* (2013, p. 2) define the field 'as the study of the collection, management, analysis, interpretation, and presentation of relational data.' In essence, a network is a collection of points, i.e., *vertices* or *nodes*, joined together in pairs by lines, i.e., *edges* (Newman, 2013). Networks are ubiquitous, ranging from neural networks capturing the connections between the neurons in the brain, to social networks mapping human interactions or trade networks representing the exchange of goods and services. Networks are at the heart of complex systems and consequently a deep understanding of the former has to be developed to fully understand the latter (Barabási, 2016).

Within network science a powerful stream of research has emerged that deals with *network controllability* (Liu *et al.*, 2011). This represents the ability to steer a dynamical system from any initial state to any desired final state within finite time using suitable inputs. The methodology builds on nonlinear dynamics and control theory (Liu *et al.*, 2011; Liu and Barabási, 2016; Kalman, 1963; Kailath, 1980). In an *Nature* article, Liu *et al.* (2011) presented analytical tools to identify the *minimum number of driver nodes* $N_D$ in an arbitrary complex directed network that, if appropriately manipulated, can offer full control over the network. Interestingly, these analytical tools are grounded on the assumption that the controllability of nonlinear systems (networks) is often structurally similar to and determined by the system's linearized dynamics (Gao *et al.*, 2014; Slotine and Li, 1991). In other words, for the detection of the minimum number of driver nodes $N_D$ in nonlinear dynamic networks, network scientists revert to the same method system dynamicists have been using in eigenvalue elasticity analysis (EEA) for a long time: approximate nonlinear dynamic systems with linearized systems near their equilibrium points (Oliva, 2016, 2015).

In this paper, we show that the network controllability framework can be applied to SD models to facilitate the discovery of model levers, i.e., effective leverage points, in the model analysis phase. More specifically, we use network controllability to identify the minimum number of driver nodes $N_D$ (variables) in SD models that is sufficient, if handled appropriately, to exert full model control according to network theory. This is relevant for policy design because the detection of high-leverage points is still an exceedingly difficult task in complex SD models. This study follows the path opened up by Moschoyiannis *et al.* (2016) and Penn *et al.* (2017) who successfully applied network controllability to fuzzy cognitive maps, a modeling field related to SD.

So how can the analytical tools of network controllability be applied to SD models in practice? Essentially, an SD model can be imagined as a web of interrelated causal factors that are assumed to give rise to the complex policy problem under study. Due to its web similarity, the *structure* of an SD model can be accurately described as a directed weighted network (weighted digraph), making it accessible to algorithmic exploration using concepts from the fields of graph theory and network science (Kampmann, 2012; Oliva, 2004; Schoenenberger *et al.*, 2015). This implies that variables and causal relationships in SD models can be translated into vertices connected by edges. Once an SD model is converted to a network (graph) representation, the application of the network controllability framework is straightforward.

Thus, we conceive the combination of SD and network controllability as a powerful formal analysis method that complements well established tools such as pathway participation metric (PPM), model structure analysis (MSA) or EEA. Figure 1 shows how the analytical tools of network controllability fit into the large scheme of formal analysis methods in SD. Importantly, while EEA methods and PPM link model structure to model behavior, MSA and network controllability are limited to characterizing model structure only. Obviously, both MSA and network controllability enable a less nuanced model analysis compared to the other two but they are clearly superior in case of qualitative model analysis since there behavioral information is absent.

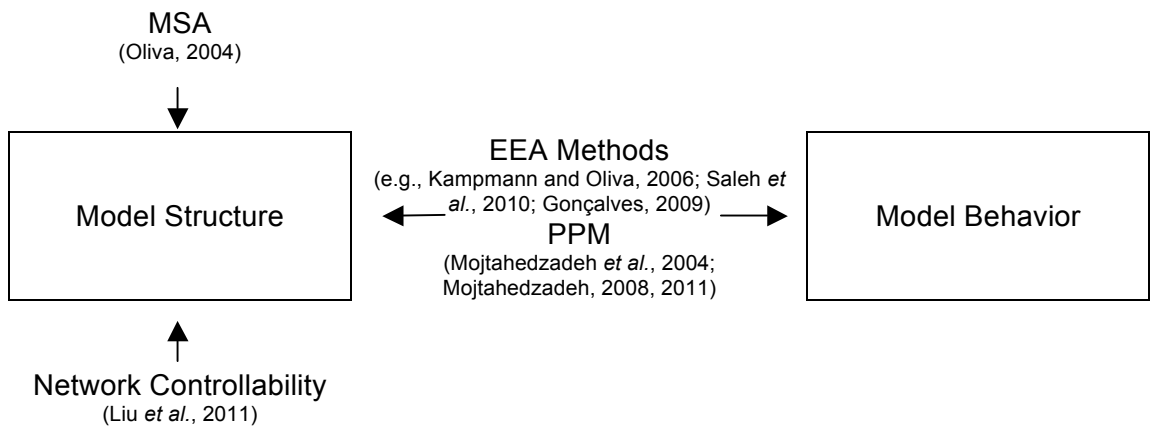

Fig. 1. Contribution of network controllability to formal model analysis in SD

The article is structured as follows: We first introduce the main concepts of network controllability and discuss the mathematical procedure to derive the minimum number of driver nodes $N_D$ in an arbitrary complex directed network that is needed to steer the entire network to any state within finite time. As typically multiple driver node configurations of size $N_D$ exist, in a next step, we describe two further node classification schemes. We then illustrate the network controllability approach using the World Dynamics model, i.e., World2 (Forrester, 1971) and discuss the potential benefits of integrating network controllability into SD for system dynamicists. We conclude by summarizing our results and provide recommendations for future research.

## Network controllability

A system is said to be controllable if we can steer it from any initial state to any desired final state in finite time (Kalman, 1963). Controllability can be easily illustrated with stick balancing, i.e., to balance a stick on a palm. From our experience, we know that this is possible, implying that the system must be controllable (Luenberger, 1979). In general controllability is a precondition of control, thus understanding the topology of the underlying network that determines a system's controllability provides numerous insights into the control principles of complex systems (Liu and Barabási, 2016). The approach considered in this article is based on the linear time-invariant control system,

$$\dot{x}(t) = \boldsymbol{A}x(t) + \boldsymbol{B}u(t) \qquad (1)$$

where $x(t)$ is a column vector representing the state of the $N$ nodes at time $t$, $\boldsymbol{A} \coloneqq (a_{ij})_{N \times N}$ is the state matrix capturing the weighted wiring diagram of the underlying network, $\boldsymbol{B} \coloneqq (b_{im})_{N \times M}$ is the input matrix identifying the nodes that are directly controlled, and $u(t)$ is an input vector. Additionally, $a_{ij}$ is the strength or weight with which node $j$ influences node $i$ where a positive (negative) $a_{ij}$ means the edge $j \to i$ is excitatory (inhibitory) and $a_{ij} = 0$ if node $j$ has no direct influence on node $i$; $b_{im}$ represents the strength of an external control signal $u_m(t)$ injected into node $i$. The linearized system in equation (1) is controllable if and only if the $N \times NM$ controllability matrix $\boldsymbol{C}$ has full rank, i.e.,

$$rank(\boldsymbol{C}) = N \qquad (2)$$

where $\boldsymbol{C} \coloneqq [\boldsymbol{B}, \boldsymbol{AB}, \boldsymbol{A}^2\boldsymbol{B}, \dots, \boldsymbol{A}^{N-1}\boldsymbol{B}]$ (Kalman, 1963). If equation (2) is satisfied, then we can find an appropriate input vector $u(t)$ to steer the system from any initial state $x(0)$ to an arbitrary final state $x(t)$, implying that the system is controllable (Liu *et al.*, 2011). From the definition of the controllability matrix $C$ it becomes clear that the network topology, captured by $A$, has a significant impact on controllability. In large networks, however, calculating $C$ is computationally demanding and often the system parameters, i.e., the elements in $A$, are not precisely known. To circumvent the latter problem, Liu *et al.* (2011) use *structural controllability* (Lin, 1974) where $A$ and $B$ are considered structured matrices, i.e., their elements are either fixed zeros or independent free parameters. The system in equation (1) is structurally controllable if we can fix the nonzero elements in $A$ and $B$ such that the resulting system satisfies equation (2). This has the advantage that we can perform the controllability test described in equation (2) even in the absence of complete knowledge of all edge weights $a_{ij}$ in the network (Liu *et al.*, 2011).

Any network is entirely controllable if we control each node individually. However, in practice this is almost always not feasible and thus we are interested in identifying the smallest subset of nodes, i.e., the minimum number of driver nodes $N_D$, that when steered by different input signals, can offer full control over the network. In other words, we want to control a network with minimal inputs (Liu and Barabási, 2016). Equation (2) will not help in finding $N_D$ because it only tells if a network is controllable or not. However, it can be shown that identifying $N_D$ is equivalent to the *maximum matching* of the network, a purely graph theoretical problem (Liu *et al.*, 2011; Lovász and Plummer, 2009). The maximum matching is the maximal set of edges in a network (graph) that do not share common nodes. A node is considered matched if there is an edge in the maximum matching set that points to it. It has been proven that we can gain full control over a directed network if and only if we directly control each unmatched node and directed paths from the input signals to all matched nodes exist (Liu *et al.*, 2011).

Thus, to fully control a directed network $G(A)$, the minimum number of driver nodes $N_D$, is

$$N_D = max\{N - |M^*|, 1\} \qquad (3)$$

where $|M^*|$ is the size of the maximum matching in $G(A)$, i.e., the number of matched nodes. Put differently, the minimum number of driver nodes $N_D$ in a network can be determined from the number of unmatched nodes $N - |M^*|$. In the limit case when all nodes are matched ($|M^*| = N$) only one input is needed to control the entire network, i.e., $N_D = 1$. A maximum matching of a directed network can be efficiently found using the Hopcroft-Karp algorithm (Hopcroft and Karp, 1973). However, as there might be multiple maximum matchings for a directed network $G(A)$, so can multiple driver node configurations exist, all of size $N_D$, that can be used for network control. For this reason, to better characterize the role of individual nodes in control, network scientists developed several node classification schemes. Jia *et al.* (2013) suggest classifying nodes according to their probability of being included in a driver node configuration. A node is

1) **critical** if that node must always be controlled to control the system, implying that it is part of all driver node configurations;
2) **intermittent** if it is a driver node in some driver node configurations but not in all;
3) **redundant** if it is never required for control, implying that it is not part of any driver node configuration.

Alternatively, Liu *et al.* (2012) introduced *control centrality* to quantify the ability of a single node in controlling an entire network. Centrality measures, i.e., tools to measure the relative importance of nodes, have a long tradition in network research. Depending on the research context, centrality measures such as degree centrality, closeness centrality, betweenness centrality, eigenvector centrality, PageRank, hub and authority centrality, and routing centrality have proven useful. Referring to these classic centrality measures, Schoenenberger and Schenker-Wicki (2014) presented a first attempt to apply them to an SD model.

In mathematical terms, the control centrality of node $i$ captures the dimension of the controllable subspace or the size of the controllable subsystem when we control node $i$ only. This can be measured with the rank of the controllability matrix $C$, defined as $rank(C)$, indicating the dimension of the controllable subspace of the linearized system in equation (1). So when we control node $i$ only, the input matrix $B$ reduces to the vector $b^i$ with a single non-zero entry, and $C$ becomes $C^i$. Similar to before, when the exact value of the edge weights is not entirely known, $A$ and $B$ are considered structured matrices. In this case, the size of the controllable subspace is measured using the generic rank, $rank_g$, of $C^i$ Johnston *et al.,* 2007; Liu *et al.,* 2012). Consequently, the control centrality of a node $i$, i.e., $C_c(i)$, is defined as

$$C_c(i) := rank_g(C^i). \qquad (4)$$

If $rank_g(C^i) = N$, then node $i$ alone can control the entire network, i.e., it can steer the network between any points in the $N$-dimensional state space in finite time. Any value of $rank_g(C^i)$ less than $N$ specifies the dimension of the subspace $i$ can control. Particularly, if $rank_g(C^i) = 1$ node $i$ can only control itself (Liu *et al.,* 2012). Equation (4) can also be normalized as follows

$$c_c(i) := \frac{C_c(i)}{N}. \qquad (5)$$

## Application of network controllability to the world dynamics model

For network controllability to be applied, the World Dynamics model (Forrester, 1971) needs to be translated into a directed network. First, we have slightly simplified Forrester's model by eliminating the lookup variables (time tabs) from the model. The resulting model contains 64 nodes and 93 edges (see Table 1.A. in the Appendix for the list of variables used). Second, we encoded the World Dynamics model in the form of its standard adjacency matrix $A$, i.e., the state matrix in equation (1). $A$ is a $N \times N$ square matrix that stores information about both the number of nodes and the exact location of all edges between them (Newman, 2013). In this case, $A$ has 4096 (64 x 64) entries where $a_{i,j} = 1$ if an edge from $i$ to $j$ exists and $a_{i,j} = 0$ otherwise.

Then, we determined i) if the network (i.e., the structure of the World Dynamics model) is controllable by checking if equation (2) is satisfied; ii) the minimum number of driver nodes $N_D$ that offers full control over the network; iii) the node assignment based on the node classification scheme introduced previously (Jia *et al.*, 2013); and iv) the control centrality $c_c(i)$ for every node in the World Dynamics model (Liu *et al.*, 2012). Finally, in a basic experimental set-up, we show that nodes with high control centrality indeed have a more substantial impact on model behavior than nodes with low control centrality. To perform the analysis we used the tools developed by Liu *et al.* (2011) and Liu *et al.* (2012). Table 1 summarizes the procedural steps for the application of network controllability to the World Dynamics model.

Table 1. Procedural steps for the application of network controllability to an SD model (here the World Dynamics model)

| Step | Description |
| --- | --- |
| 1 | Preprocess the SD model: |
|  | Codify the SD model into its standard adjacency matrix $A$. To simplify the coding, in a manual step, variables with no real meaning for the complex policy problem under study, i.e., lookup variables or time constants, can be omitted. |
| 2 | Use the standard adjacency matrix $A$ derived in step (1) as an input for the controllability analysis. The analysis can be done using the following C++ code packages: 'ControllabilityAnalysis' (Liu *et al.*, 2011) and 'CalControlCentrality' (Liu *et al.*, 2012). The code is available under https://scholar.harvard.edu/yyl/code [31.05.2017]; creator permission is necessary. |
| 3 | Perform the analysis in order to answer the following questions: |
|  | - Is the structure of the SD model (network) controllable at all? |
|  | - How many and which variables are sufficient to exert full control over the structure of the SD model given it is controllable? |
|  | - How important are individual variables (nodes) in controlling the structure of the SD model? |

The analysis yielded that the network under study is indeed controllable. The minimum number of driver nodes $N_D$ equals 34 implying that it is enough to control only 53% of all nodes, i.e., $\frac{N_D}{N} * 100$, to steer the entire network to any point in the *N*-dimensional state space. Figure 2 displays all nodes, i.e., variables, in the World Dynamics model and their classification into critical, intermittent, and redundant nodes is highlighted using different shapes. To avoid a 'crowded' Figure 2, we had to abbreviate the node names. An exhaustive variable list with both full names and abbreviations can be found in the Appendix. All critical nodes in the World Dynamics model, highlighted as shaded squares, are parameters meaning that they belong to all driver node configurations (of size $N_D$). This seems intuitive to system dynamicists, because it is the parameters that are their primary target when it comes to policy analysis. From a network controllability perspective, this result is not surprising since all nodes having no incoming links, i.e., the exogenous variables (parameters), must be directly controlled (Ruths *et al.,* 2014).

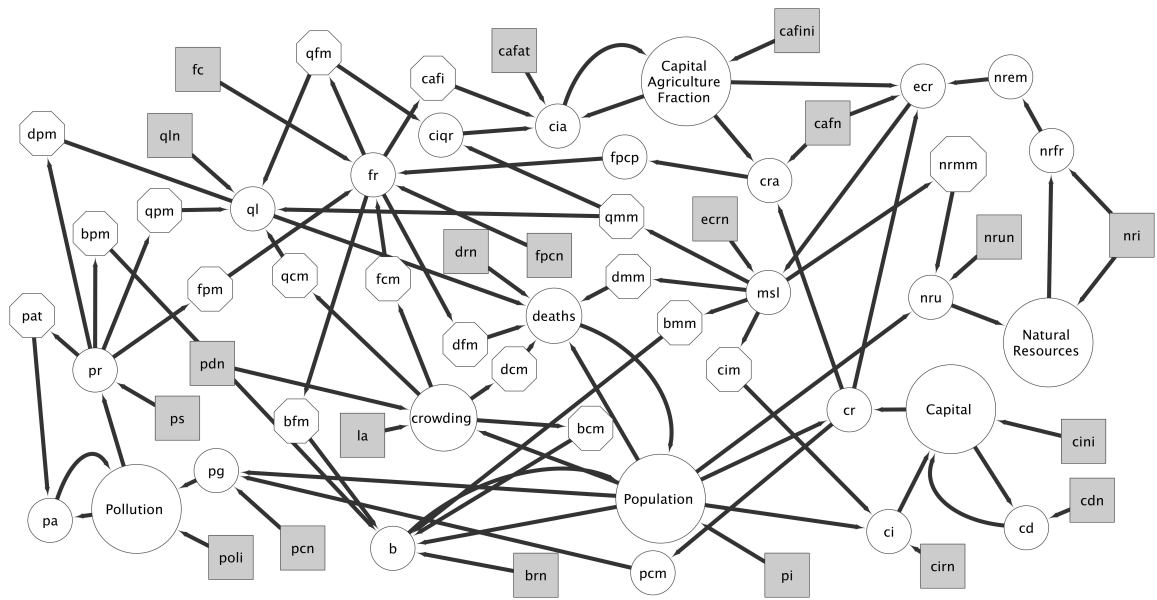

Fig. 2. Classification of nodes based on their roles in control in the World Dynamics model

The intermittent nodes in the World Dynamics model, highlighted as hexagons, are a subset of all the variables (auxiliaries) in the model. Interestingly, these variables are, with two exceptions, i.e., *pollution absorption time* (pat) and *capital agriculture fraction indicated* (cafi), multipliers in the World Dynamics model. Therefore, from a purely structural viewpoint, these multipliers seem highly relevant in controlling the World Dynamics model. Finally, the redundant nodes in the World Dynamics model, highlighted as circles, comprise all stock and flow variables, and so they do not have to be directly manipulated to control the World Dynamics model. This is consistent with SD practice where stocks cannot be directly controlled but only through their flows which in turn are steered by parameters.

Now we dive deeper into the node classification by analyzing the control capacity of individual nodes. As discussed earlier, the control centrality corresponds directly to our intuition of how powerful a single node is (or groups of nodes are) in controlling the whole network (Liu *et al.*, 2012). Table 2 shows both the 7 nodes with the highest normalized control centrality $c_c(i)$, all attaining the same score, and the 4 nodes with the lowest $c_c(i)$ in the World Dynamics model. In the model, the range of $c_c(i)$ lies between [0.02, 0.44]. We chose to display only the 4 lowest scoring nodes because they have a significantly lower $c_c(i)$ than all the other nodes in the sample (see Table 1.A. in the Appendix). The top 7 nodes all achieve a $c_c(i)$ of 0.44 and consist of five parameters and two variables (auxiliaries). In particular, the parameters might serve as effective leverage points in the World Dynamics model.

Table 2. Nodes with high and low influence on the World Dynamics model.

| Node name | Abbreviation | Type | Classification (Jia et al., 2013) | $c_c(i)$ (Liu et al., 2012) |
|---|---|---|---|---|
| *Most influential nodes with respect to normalized control centrality $c_c(i)$* | | | | |
| capital.depreciation.normal | cdn | P | critical | 0.44 |
| capital.investment.rate.normal | cirn | P | critical | 0.44 |
| land.area | la | P | critical | 0.44 |
| natural.resource.utilization.normal | nrun | P | critical | 0.44 |
| population.density.normal | pdn | P | critical | 0.44 |
| capital.investment.multiplier | cim | V | intermittent | 0.44 |
| nat.res.matl.multiplier | nrmm | V | intermittent | 0.44 |
| *Least influential nodes with respect to normalized control centrality $c_c(i)$* | | | | |
| quality.of.life | ql | V | redundant | 0.02 |
| quality.pollution.multiplier | qpm | V | intermittent | 0.03 |
| quality crowding multiplier | qcm | V | intermittent | 0.03 |
| quality.of.life.normal | qln | P | critical | 0.03 |

P: Parameters; V: Variables; $c_c(i)$: Normalized control centrality of node i

Now we test if the nodes with a high normalized control centrality, $c_c(i)$, really have a more substantial impact on the behavior of the World2 model than the ones with a low $c_c(i)$. To check this, we performed the following test:

1) We randomly picked 4 nodes from the sample of the most influential nodes (sample size = 7); and compared them against the 4 lowest scoring nodes.

2) In the sense of policy experiments, we separately increased the value of all 8 nodes by 10 % and ran 8 different simulations, i.e., in every simulation only *one* variable is changed.

3) We assessed the impact these changes have on the 5 stocks—*Population*, *Capital*, *Capital Agriculture Fraction*, *Pollution*, and *Natural Resources*—by visual inspection only. As a reference curve, a base run according to Forrester's (1971) original model parameterization is executed.

Figure 3 shows the impact of a 10 % increase of both the 4 influential nodes, i.e., the ones with the highest $c_c(i)$, and the 4 ineffective nodes, i.e., the ones with the least $c_c(i)$, on the key stock variable in the World2 model: *Population*.

Most notable, individually increasing the 4 least scoring nodes by 10 % has no impact at all on *Population* and the other stock variables (not shown in Figure 3). In contrast, raising the 4 highest scoring nodes by 10 % has a significant impact on the trajectory of *Population* and the other stock variables. Particularly, increasing the parameter *capital investment rate normal* by 10 % not only changes the maximum and final equilibrium point of the trajectory but also its general shape (mode of behavior), i.e., from overshoot and decay to damped oscillation. In conclusion, the test provides a strong indication that high scoring nodes have much more influence on the behavior of the World2 model than low scoring nodes.

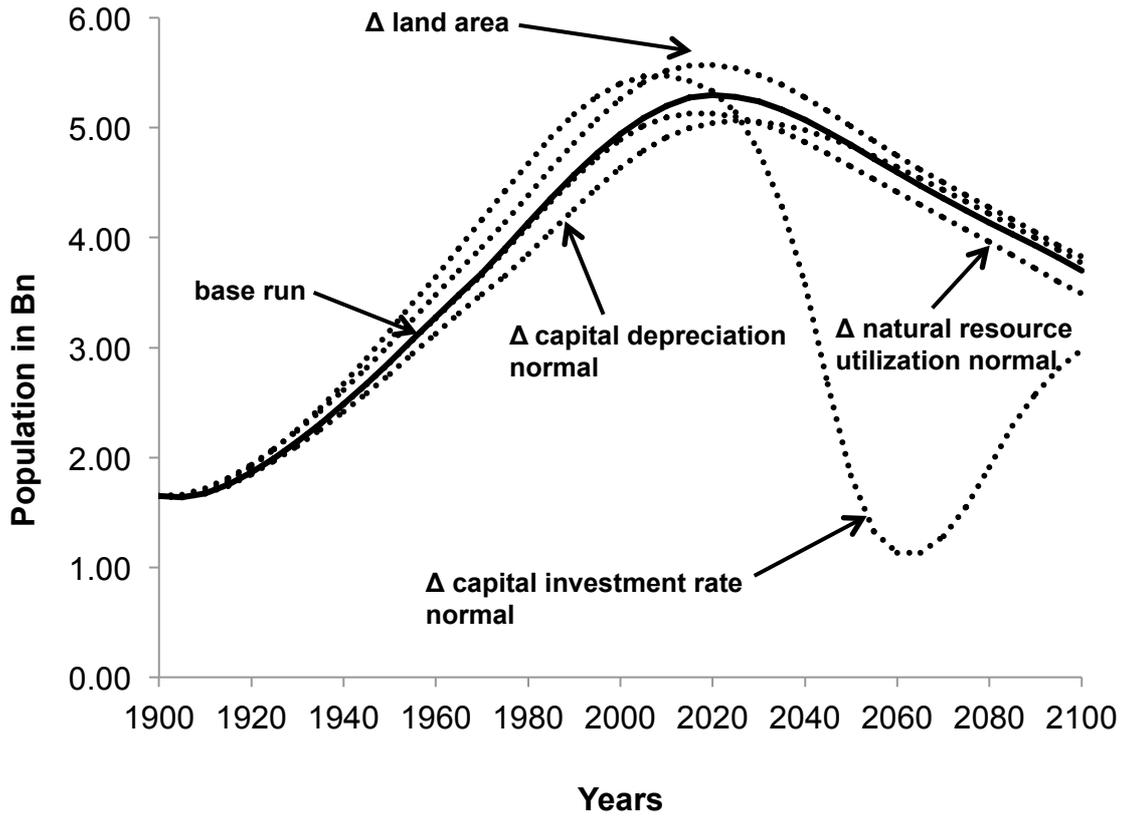

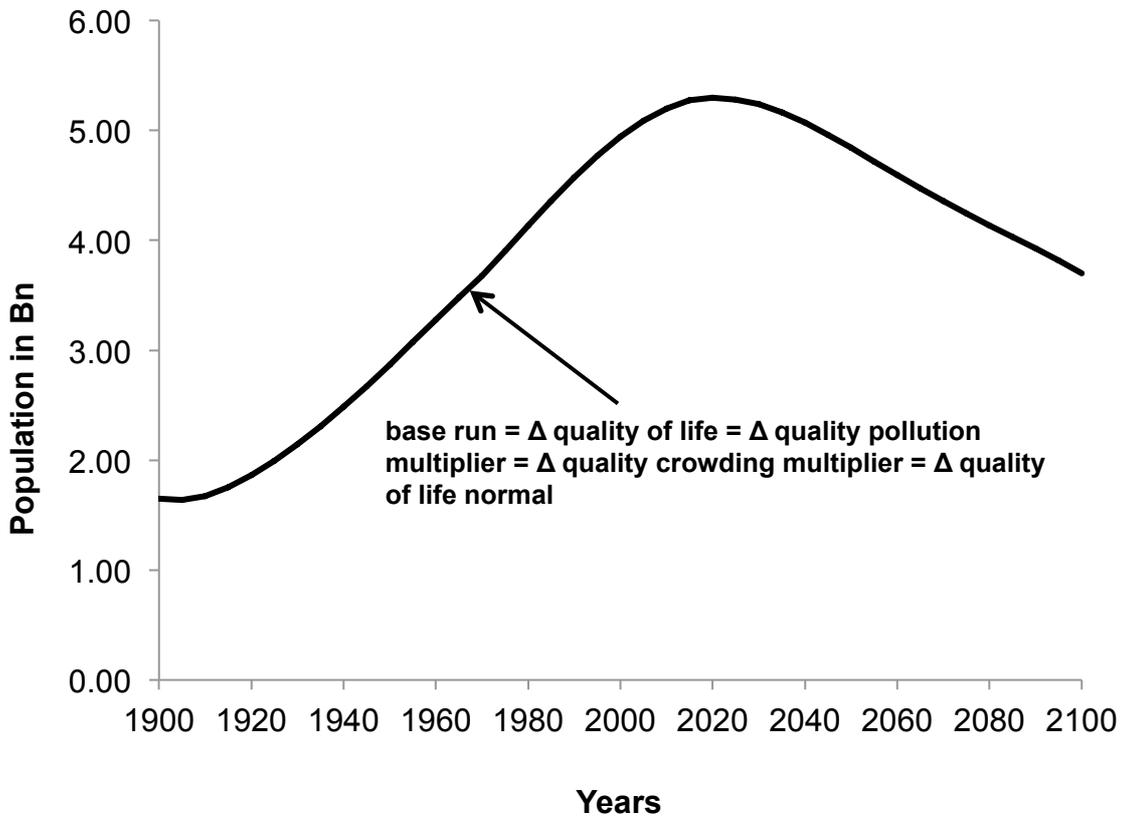

Fig. 3. Impact of a 10 % increase in the 4 top scoring nodes (upper diagram) and in the 4 least scoring nodes (lower diagram) on *Population*

## Integration of network controllability into system dynamics

We believe that network controllability is a good complement to the formal analysis techniques in SD (see Figure 1). Particularly, we see a significant synergistic potential with MSA (Oliva, 2004) which mainly focuses on feedback structures. Based on a purely structural comprehension of SD models, Oliva (2004) is able to derive the hierarchy of feedback loops in models. In contrast, network controllability concentrates on single nodes and their role in the control of directed networks. In principle, we see two possible options for integrating network controllability into the SD process:

1)  Integration of network controllability into model analysis (focus of this article):

    Alongside other well-established formal analysis techniques, network controllability might serve as a first screening tool of complex SD models for the purpose of identifying leverage points (policy design) within them.

2)  Integration of network controllability into model building:

    Network controllability has the potential to guide the model building process. It is probably most useful when small models are expanded to medium sized ones or when qualitative conceptual maps are transformed into working simulation models. This is because network controllability helps to focus model building on variables that are crucial to the complex policy problem under study. In this context, network controllability might support system dynamicists in defining model regions that are worthwhile to expand or in defining key variables that need to be parameterized for a quantitative simulation model.

Figure 4 illustrates the standard SD process and its possible interfaces to network controllability. In this paper, we illustrated the potential of integrating network controllability into model analysis. This can serve as a preliminary screening tool to identify potential leverage points in SD models as we have demonstrated on the basis of the World Dynamics model. In other words, such an additional structural analysis can assist system dynamicists in designing alternative policies and structures (step 4 in Figure 4).

Traditionally, these alternative policies come from intuitive insights generated in the preceding steps of the SD process, from the experience of the modeler, from people operating in the system of interest, or by an exhaustive automatic testing of parameter changes (Forrester, 1994). Consequently, the development of effective alternative policies is difficult, especially in large models, and so a strategy for preliminary determination of candidate nodes for policy design is very helpful.

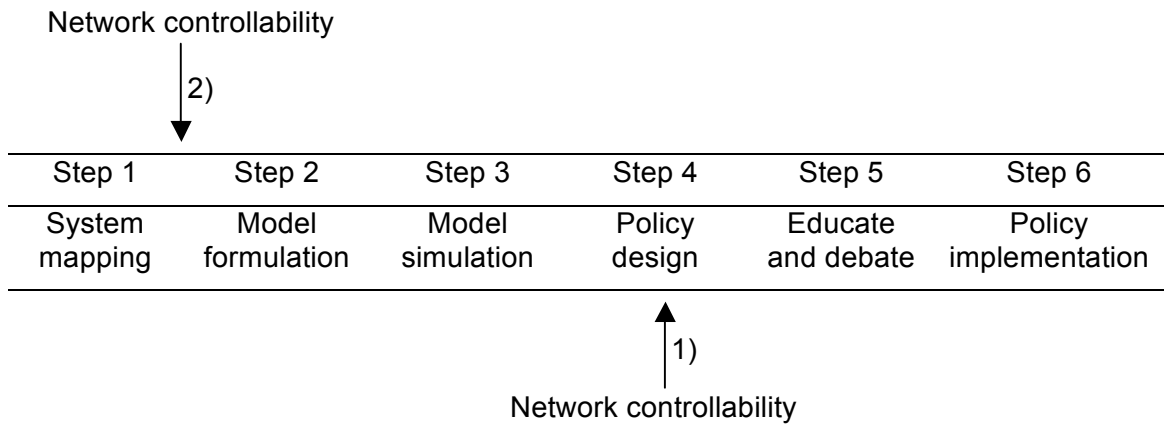

Fig. 4. SD process, based on Forrester (1994), and its interfaces to network controllability

## Discussion and conclusions

In this article, we argue for an integration of network controllability into the SD process to enhance the current toolset of system dynamicists in formal model analysis. More specifically, we conceive of network controllability as a powerful complementary tool to MSA in exploring the structure of SD models. In contrast to MSA, which deals with feedback complexity, network controllability focuses on single nodes and their role in the control of directed networks. Therefore, network controllability, might be most valuable as a preliminary screening tool of complex SD models to detect potential leverage points within them (see step 4 in Figure 4). Every modeler is confronted with two key challenges: *how* to best represent or model the system, and *where* to change the system to generate more favorable system outcomes. We believe that network controllability can help modelers address the latter problem by providing such a screening tool.

Merging SD with network controllability is a new approach and has limitations that prescribe future research avenues. First, network controllability applied to SD offers a less nuanced analysis than EEA methods or PPM because it is limited to exploring model structure only. It is clear that system dynamicists are most interested in system behavior and not in structure *per se*. However, one of the key pillars of SD emphasizes that system behavior arises from underlying system structure (Oliva, 2004; Meadows, 1989). Second, while network controllability provides information about which nodes modelers must tackle for full network control, it does not say how and how much nodes have to be changed. As a consequence, nodes might have to be changed by so much that it is infeasible to implement this change in practice. Third, so far we have only shown the effectiveness of control central nodes in steering model behavior in a basic experimental setting. Thus, future research should be directed towards a systematic investigation of the effect of control central nodes on model behavior by evaluating multiple SD models. Fourth, network controllability builds on the strong assumption that linearizing a nonlinear dynamic system near its equilibrium point is a reasonable procedure. In EEA, so far, this assumption has served well but it is unclear if this holds for the combination of network controllability with SD as well. It is certain, however, when model nonlinearities begin to be dominant determinants of model behavior, the value of linear analysis is limited (Eberlein, 1989).

Finally, a modeler might often be interested to tune (refine) a model sector only and thus to concentrate the analysis on one specific model region. The methodology we have described so far can only accommodate such cases when the sub-model of interest can be treated as an independent part of the rest of the model. When this is the case, one can reduce the standard adjacency matrix $A$ to an $L$ dimensional matrix, where $L$ represents the number of variables in the sub-model, and use this as an input to the controllability analysis. However, such cases are rarely encountered in practice and thus an extension of the current methodology to account for controlling sub-networks while using the entire network structure might provide an important avenue for future research.

## Acknowledgements

The work of R.T. was supported by the University Research Priority Program "Social Networks" of the University of Zurich.

# Appendix

Table 1.A. Variable list with full names, abbreviations, classification (Jia *et al.*, 2013), and $c_c(i)$ scores. Variables are ordered alphabetically.

| Variable | Abbreviation | Type | Classification | $c_c(i)$ |
|---|---|---|---|---|
| birth rate normal | brn | P | critical | 0,41 |
| births | b | F | redundant | 0,39 |
| births crowding multiplier | bcm | V | intermittent | 0,41 |
| births food multiplier | bfm | V | intermittent | 0,41 |
| births material multiplier | bmm | V | intermittent | 0,41 |
| births pollution multiplier | bpm | V | intermittent | 0,41 |
| Capital | Capital | S | redundant | 0,41 |
| Capital Agriculture Fraction | Capital Agriculture Fraction | S | redundant | 0,38 |
| capital agriculture fraction adjustment time | cafat | P | critical | 0,39 |
| capital agriculture fraction indicated | cafi | V | intermittent | 0,39 |
| capital agriculture fraction initial | cafini | P | critical | 0,39 |
| capital agriculture fraction normal | cafn | P | critical | 0,39 |
| capital depreciation | cd | F | redundant | 0,42 |
| capital depreciation normal | cdn | P | critical | 0,44 |
| capital initial | cini | P | critical | 0,42 |
| capital investment | ci | F | redundant | 0,42 |
| capital investment from quality ratio | ciqr | V | redundant | 0,39 |
| capital investment in agriculture | cia | F | redundant | 0,38 |
| capital investment multiplier | cim | V | intermittent | 0,44 |
| capital investment rate normal | cirn | P | critical | 0,44 |
| capital ratio | cr | V | redundant | 0,42 |
| capital ratio agriculture | cra | V | redundant | 0,38 |
| crowding | crowding | V | redundant | 0,42 |
| death rate normal | drn | P | critical | 0,41 |
| deaths | deaths | F | redundant | 0,39 |
| deaths crowding multiplier | dcm | V | intermittent | 0,41 |
| deaths food multiplier | dfm | V | intermittent | 0,41 |
| deaths material multiplier | dmm | V | intermittent | 0,41 |
| deaths pollution multiplier | dpm | V | intermittent | 0,41 |
| effective capital ratio | ecr | V | redundant | 0,38 |
| effective capital ratio normal | ecrn | P | critical | 0,39 |
| food coefficient | fc | P | critical | 0,39 |
| food crowding multiplier | fcm | V | intermittent | 0,39 |
| food per capita normal | fpcn | P | critical | 0,39 |
| food per capita potential | fpcp | V | redundant | 0,38 |
| food pollution multiplier | fpm | V | intermittent | 0,39 |
| food ratio | fr | V | redundant | 0,38 |
| land area | la | P | critical | 0,44 |
| material standard of living | msl | V | redundant | 0,38 |
| nat res matl multiplier | nrmm | V | intermittent | 0,44 |
| natural resource extraction multiplier | nrem | V | redundant | 0,38 |
| natural resource fraction remaining | nrfr | V | redundant | 0,39 |
| natural resource utilization | nru | F | redundant | 0,42 |
| natural resource utilization normal | nrun | P | critical | 0,44 |
| Natural Resources | Natural Resources | S | redundant | 0,41 |
| natural resources initial | nri | P | critical | 0,42 |
| Pollution | Pollution | S | redundant | 0,38 |
| pollution absorption | pa | F | redundant | 0,39 |
| pollution absorption time | pat | V | intermittent | 0,41 |
| pollution capital multiplier | pcm | V | redundant | 0,41 |
| pollution generation | pg | F | redundant | 0,39 |
| pollution initial | poli | P | critical | 0,39 |
| pollution per capita normal | pcn | P | critical | 0,41 |
| pollution ratio | pr | V | redundant | 0,39 |
| pollution standard | ps | P | critical | 0,41 |
| Population | Population | S | redundant | 0,38 |
| population density normal | pdn | P | critical | 0,44 |
| population initial | pi | P | critical | 0,39 |
| quality crowding multiplier | qcm | V | intermittent | 0,03 |
| quality food multiplier | qfm | V | intermittent | 0,41 |
| quality material multiplier | qmm | V | intermittent | 0,41 |
| quality of life | ql | V | redundant | 0,02 |
| quality of life normal | qln | P | critical | 0,03 |
| quality pollution multiplier | qpm | V | intermittent | 0,03 |

P: Parameter; V: Variable; S: Stock; F: Flow